\documentclass[aps,prb,twocolumn,showpacs,amsmath,amssymb,unsortedaddress]{revtex4-1}
\usepackage{graphicx}
\usepackage{longtable}
\usepackage{ulem}
\usepackage{color}

\begin{document}
\title{Evidence for structural transition in crystalline tantalum pentoxide films grown by RF magnetron sputtering}

\author{Israel Perez} 
\email[Contact Author: ]{cooguion@yahoo.com}
\affiliation{National Council of Science and Technology (CONACYT)-Institute of Engineering and Technology, Universidad Aut\'onoma de Ciudad Ju\'arez, Av. del Charro 450 Col. Romero Partido, C.P. 32310, Ju\'arez, Chihuahua, M\'exico}
\author{Jos\'e Luis Enr\'iquez Carrejo} 
\affiliation{Institute of Engineering and Technology, Universidad Aut\'onoma de Ciudad Ju\'arez, Av. del Charro 450 Col. Romero Partido, C.P. 32310, Ju\'arez, Chihuahua, M\'exico}
\author{V\'ictor Sosa} 
\affiliation{Applied Physics Department, CINVESTAV Unidad M\'erida, km 6 Ant. Carretera a Progreso, A.P. 73, C.P. 97310 M\'erida, Yucat\'an, M\'exico}
\author{Fidel Gamboa Perera}
\affiliation{{\it Applied Physics Department, CINVESTAV Unidad M\'erida, km 6 Ant. Carretera a Progreso, A.P. 73, C.P. 97310 M\'erida, Yucat\'an, M\'exico}}
\author{Jos\'e Rurik Farias Mancillas}
\affiliation{{\it Institute of Engineering and Technology, Universidad Aut\'onoma de Ciudad Ju\'arez, Av. del Charro 450 Col. Romero Partido, C.P. 32310, Ju\'arez, Chihuahua, M\'exico}}
\author{Jos\'e Trinidad Elizalde Galindo} 
\affiliation{Institute of Engineering and Technology, Universidad Aut\'onoma de Ciudad Ju\'arez, Av. del Charro 450 Col. Romero Partido, C.P. 32310, Ju\'arez, Chihuahua, M\'exico}
\author{Carlos Iv\'an Rodr\'iguez Rodr\'iguez} 
\affiliation{Universidad Tecnol\'ogica de Ciudad Ju\'arez, Av. Universidad Tecnol\'ogica No. 3051, Col. Lote Bravo II, C.P. 32695, Ju\'arez, Chihuahua, M\'exico}
\date{\today}

\begin{abstract}
We investigate the effect of annealing temperature on the crystalline structure and physical properties of tantalum-pentoxide films grown by radio frequency magnetron sputtering. For this purpose, several tantalum films were deposited and the Ta$_2$O$_5$ crystalline phase was induced by exposing the samples to heat treatments in air in the temperature range from (575 to 1000)$^\circ$C. Coating characterization was performed using X-ray diffraction, scanning electron microscopy, Raman spectroscopy and UV-VIS spectroscopy. By X-ray diffraction analysis we found that a hexagonal Ta$_2$O$_5$ phase generates at temperatures above $675^\circ$C. As the annealing temperature raises, we observe peak sharpening and new peaks in the corresponding diffraction patterns indicating a possible structural transition from hexagonal to orthorhombic. The microstructure of the films starts with flake-like structures formed on the surface and evolves, as the temperature is further increased, to round grains. We  found out that, according to the features exhibited in the corresponding spectra, Raman spectroscopy can be sensitive enough to discriminate between the orthorhombic and hexagonal phases of Ta$_2$O$_5$. Finally, as the films crystallize the magnitude of the optical band gap increases from 2.4 eV to the typical reported value of 3.8 eV.

\end{abstract}
\maketitle
\section{Introduction}
Ta$_2$O$_5$ is a widely known material used in the electronic and chemical industries for the production of insulators, capacitors, gas detectors, catalysts and proton conductors \cite{tkaga91a,kwkwon96,cchaneliere98a}. It is a semiconductor with a wide optical band gap $E_g$ ($\sim$4.0 eV) and high dielectric constant that results in exceptional optical properties such as a high refraction index ($n=2.18$ at $\lambda=550 \;\text{nm}$)\cite{cchaneliere98b,rhdennard74a,sshibata96a,eatanassova99a}. Due to these properties, it also finds common usage in solar cells and optical waveguides \cite{jdkruschwitz97a}. Ta$_2$O$_5$ nucleates in either amorphous or crystalline phases. The latter has received much attention in recent years due to an enhancement of the electrical properties ($\epsilon_r>25$) which opens the door for additional applications \cite{eatanassova99a,jdkruschwitz97a,cchaneliere99a}.

The study of the crystalline phase of Ta$_2$O$_5$ in the form of films is of great importance for the electronics industry. Most methodologies used in the fabrication of Ta$_2$O$_5$ coatings involve post-deposition heat treatments above 600$^\circ$C \cite{tdimitrova01a,sjjwu09a,dcristea13a}. The crystallinity of the phase depends on several factors such as preparation method, annealing temperature $T_{ann}$, film thickness $t$ and substrate temperature $T_s$. Earlier research showed evidence for the existence of several polymorphs for temperatures below 1320$^\circ$C, the so-called low temperature phases denoted as $L-$Ta$_2$O$_5$ \cite{spgarg96a,ktjacob09a}. Most scientists agree that the system crystallizes in either orthorhombic or hexagonal structures although the exact symmetry is still under scrupulous investigation \cite{spwalton16a}. This issue becomes critical in the production of thin and ultra-thin films. Usually, Si substrates are used for the deposition of coatings, however, their structural characterization represents a challenge due to the small dimensions of the samples. Frequently, XRD (X-ray diffraction) measurements reveal the presence of diffraction peaks coming from the substrate that usually overlap with the peaks generated from the film which evidently precludes a proper phase identification \cite{jykim14a,pshang13a}. Another difficulty that arises during film growth is the formation of an oxide interface layer between the substrate and the film of interest \cite{tdimitrova01a,chan94a,amuto94a, aphuang05a,mzhu06a}. This layer can be as thick as 5 nm and hence generates additional diffraction peaks that may overlap with those generated by the film, thus worsening the issue of phase indexing. Furthermore, the relatively high annealing temperature may cause problems during semiconductor manufacturing and thus exclude Ta$_2$O$_5$ films from electronic applications \cite{svjagadeesh10a}. It is therefore worth investigating the effects of the annealing temperature on the evolution of the crystallinity and physical properties of Ta$_2$O$_5$ films.

To shed some light on these matters, several Ta films were deposited on Si substrates by RF magnetron sputtering and then induced the crystalline phase of Ta$_2$O$_5$ by exposing the films to different heat treatments. We then studied the impact of the annealing temperature on the crystalline structure, the morphology, the vibrational modes, and the optical band gap. During film deposition we did not use oxygen nor heated the substrate. Instead, with the aim of avoiding the effects of both the substrate and the oxide interfacial layer, we grew various thick films and exposed them to an external heat treatment in air. For the study of the crystalline and morphological evolution of the coatings we used XRD and scanning electron microscopy (SEM), respectively. The active vibrational modes were analyzed with Raman spectroscopy and the optical properties such as the band gap and absorption coefficient were measured with ultraviolet-visible spectroscopy (UV-VIS).
 
\section{Experimental}
\subsection{Film growth and annealing}
Six thick amorphous Ta films were grown on 5 mm $\times$ 7 mm--Si(100) substrates by the RF magnetron sputtering technique. The films were intentionally deposited at room temperature from a Ta target (99.95\% purity) with a substrate-target distance of 13 cm. Before deposition, the vacuum chamber was initially evacuated at a base pressure of $5.0 \times 10^{-5}$ Torr. Immediately, argon gas was flushed into the chamber up to a working pressure of ($2.0\pm 0.2)\times10^{-2}$ Torr. To eliminate the oxide layer on the target surface, a 5-min pre-sputtering process at 60 W was performed. Subsequently, for the deposition of the coatings, the power was increased up to 120 W. To ensure both the atomic homogeneity of the films and the same film thickness all over the substrate surface, the substrate was situated on a rotatory base spinning at a speed of 0.2 rpm. After deposition, five films were exposed to heat treatments in air for 1 h at different temperatures for each film using a Thermo Scientific Thermolyne cylindrical furnace (model F21135). These films were labeled as F575, F675, F775, F875, and F1000; the numbers referring to the annealing temperature. We kept as reference the remaining as-deposited film which was not exposed to the heat treatment and was labeled F0. The film parameters such as annealing temperature, film thickness $t$, deposition rate $D_r$, etc. are tabulated in Table \ref{details}.

\subsection{Characterization}
The crystalline structure of the samples was studied by XRD using a Siemens diffractometer model D-5000 with Cu $K_{\alpha}$ radiation ($\lambda=1.5406$ \AA). XRD patterns were measured at steps of 0.02$^\circ$ with a time per step of 3 s in a Bragg-Bretano configuration and with operating parameters of 34 kV and 25 $\mu$A. The analyses of morphology, particle size, and microstructure were carried out by scanning electron microscopy (SEM) in a field emission microscope JSM7000F. Sample stoichiometry was calculated from the spectra of the Ta4f and O1s bands using X-ray photoelectron spectroscopy (XPS). For this purpose we used a Thermo Scientific K-Alpha XPS spectrometer with an Al K$_{\alpha}$ X-ray source set to 12 kV and 40 W, and sweeping energy at steps of 0.1 eV. The beam spot has a diameter of 400 $\mu$m and makes an angle relative to the sample of 30$^\circ$. Raman spectroscopy measurements were realized on a WITec alpha 300 R confocal Raman system with a Nd:YAG source ($\lambda= 532$ nm) and maximum output power of 50 mW. Film thickness was determined in situ by a QCM (quartz crystal monitor) and ex situ by SEM and within the limits of error both measurements were in agreement with each other.  Lastly, the optical properties were assessed with the help of an UV-VIS spectrometer Perkin-Elmer, Alpha 25 in the reflection mode used in diffuse reflection configuration. 

\section{Results and discussion}

\subsection{Surface morphology}
\begin{figure}[t!]
\begin{center}
\includegraphics[width=4.0cm]{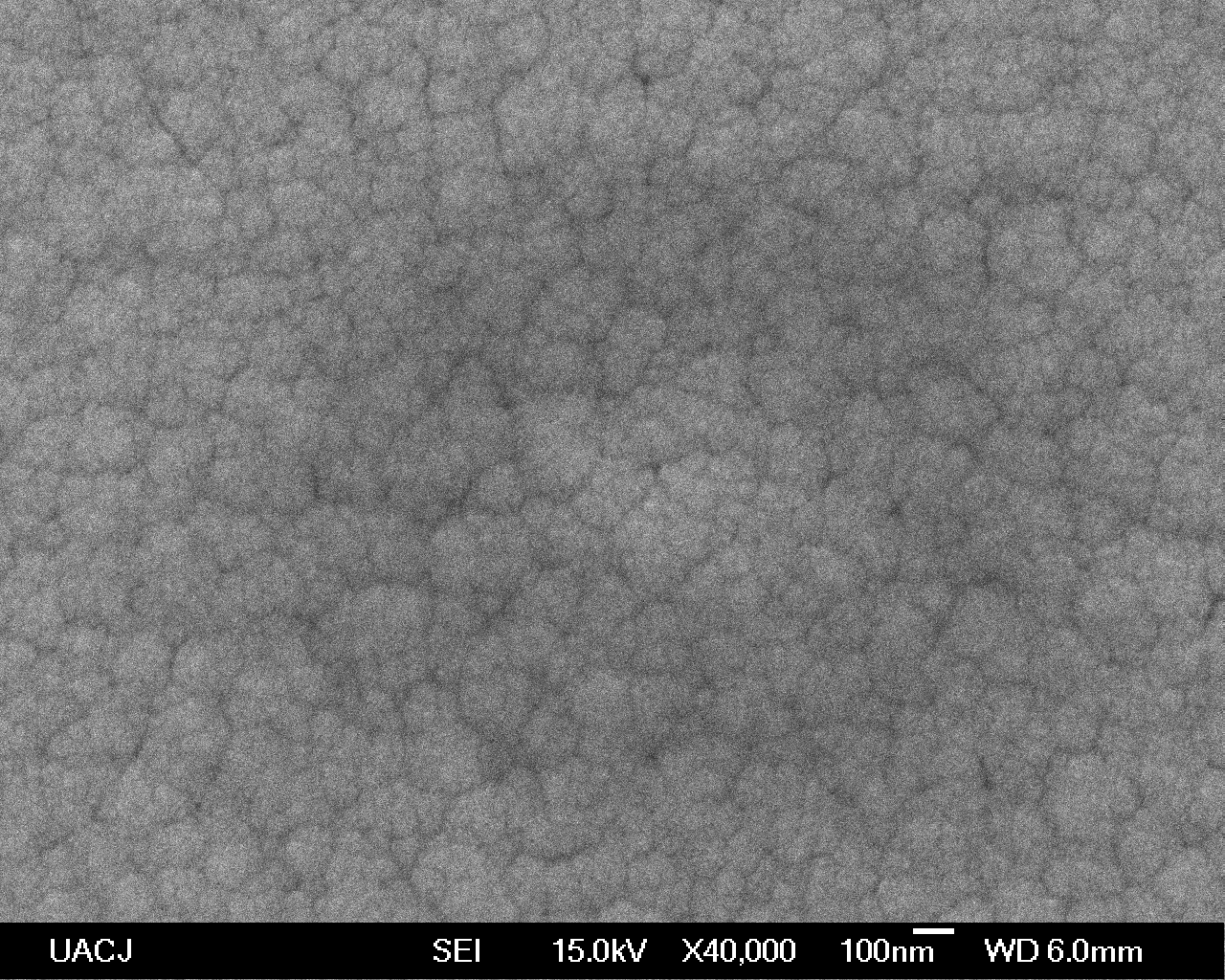} \includegraphics[width=4.0cm]{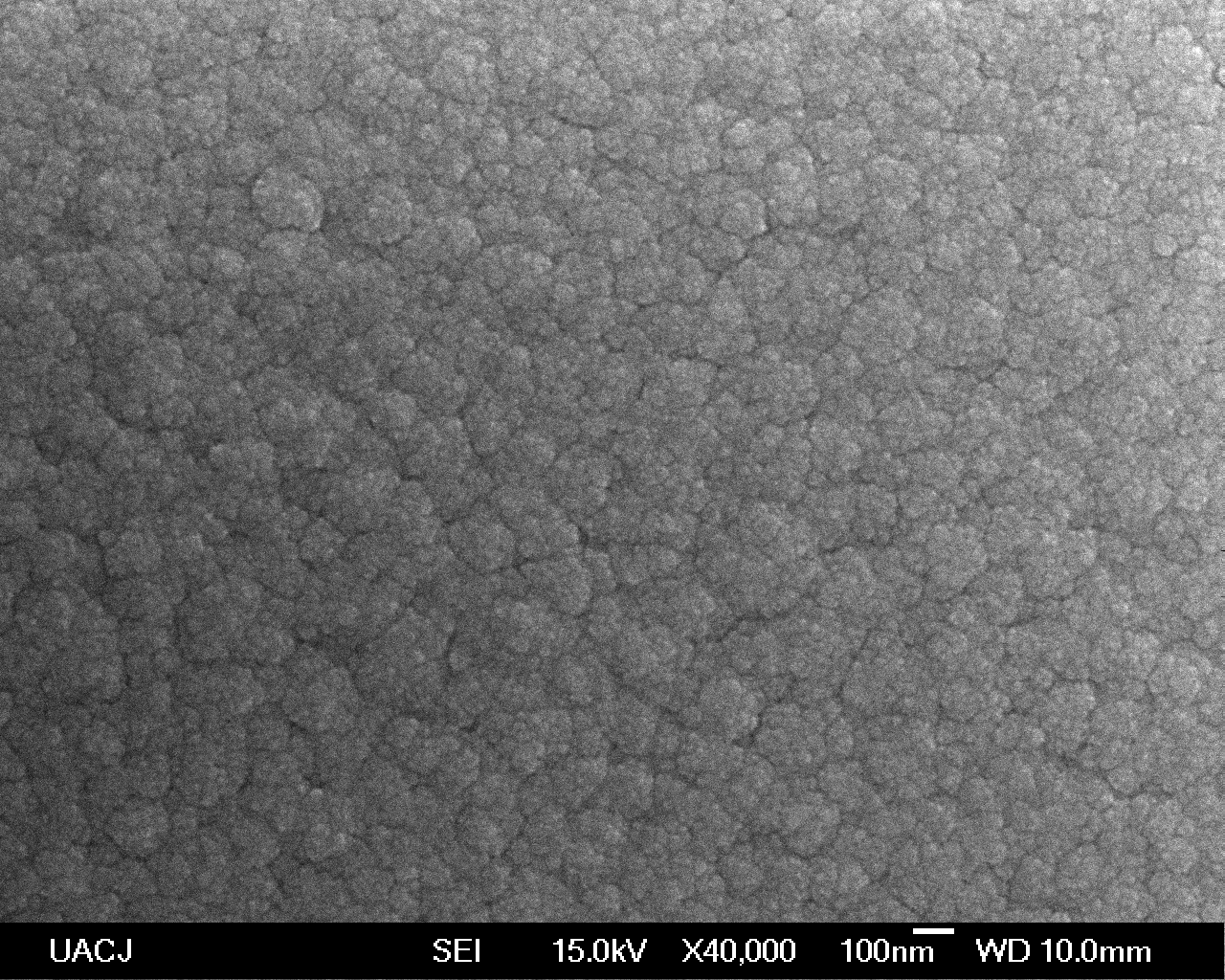}
\includegraphics[width=4cm]{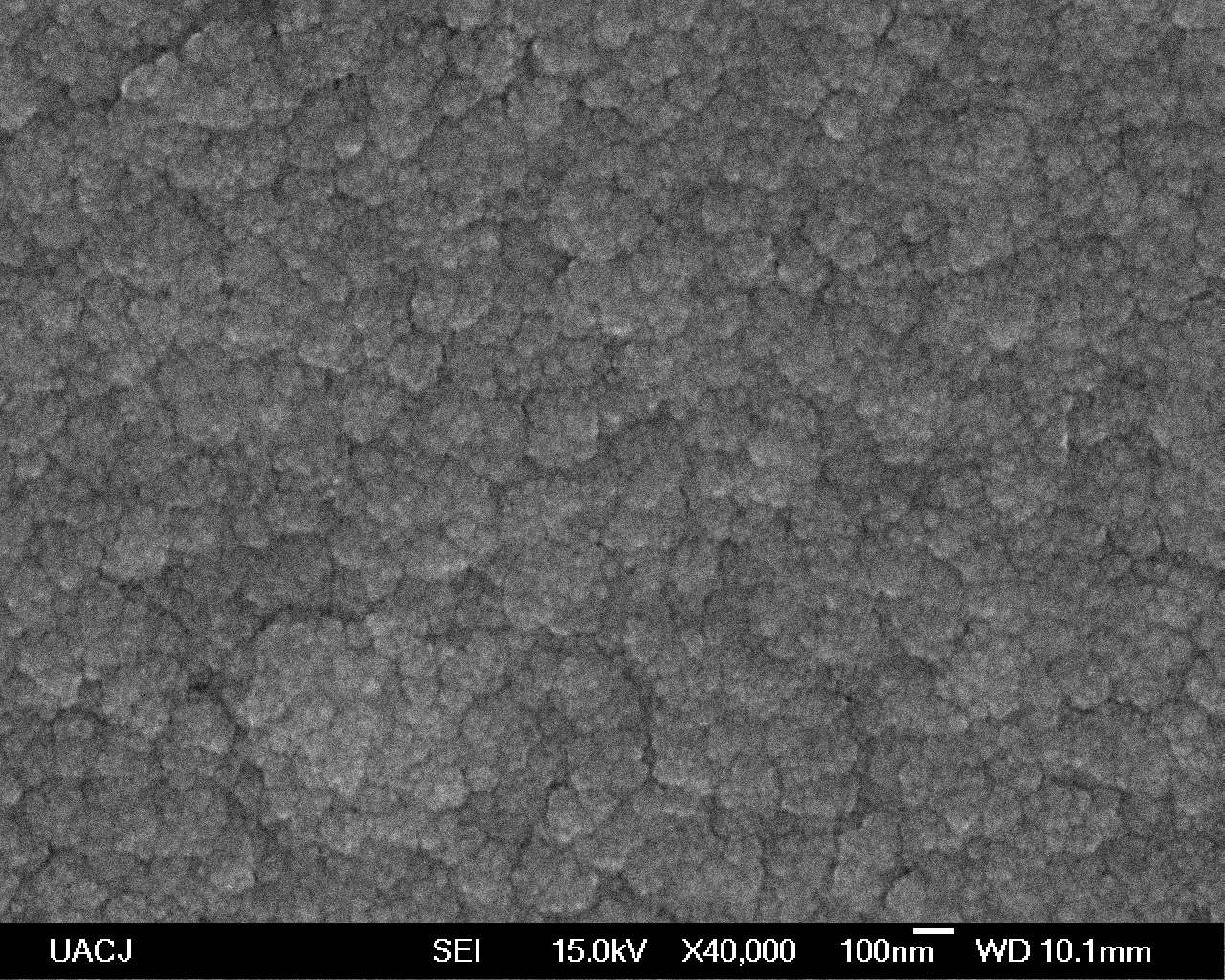} \includegraphics[width=4cm]{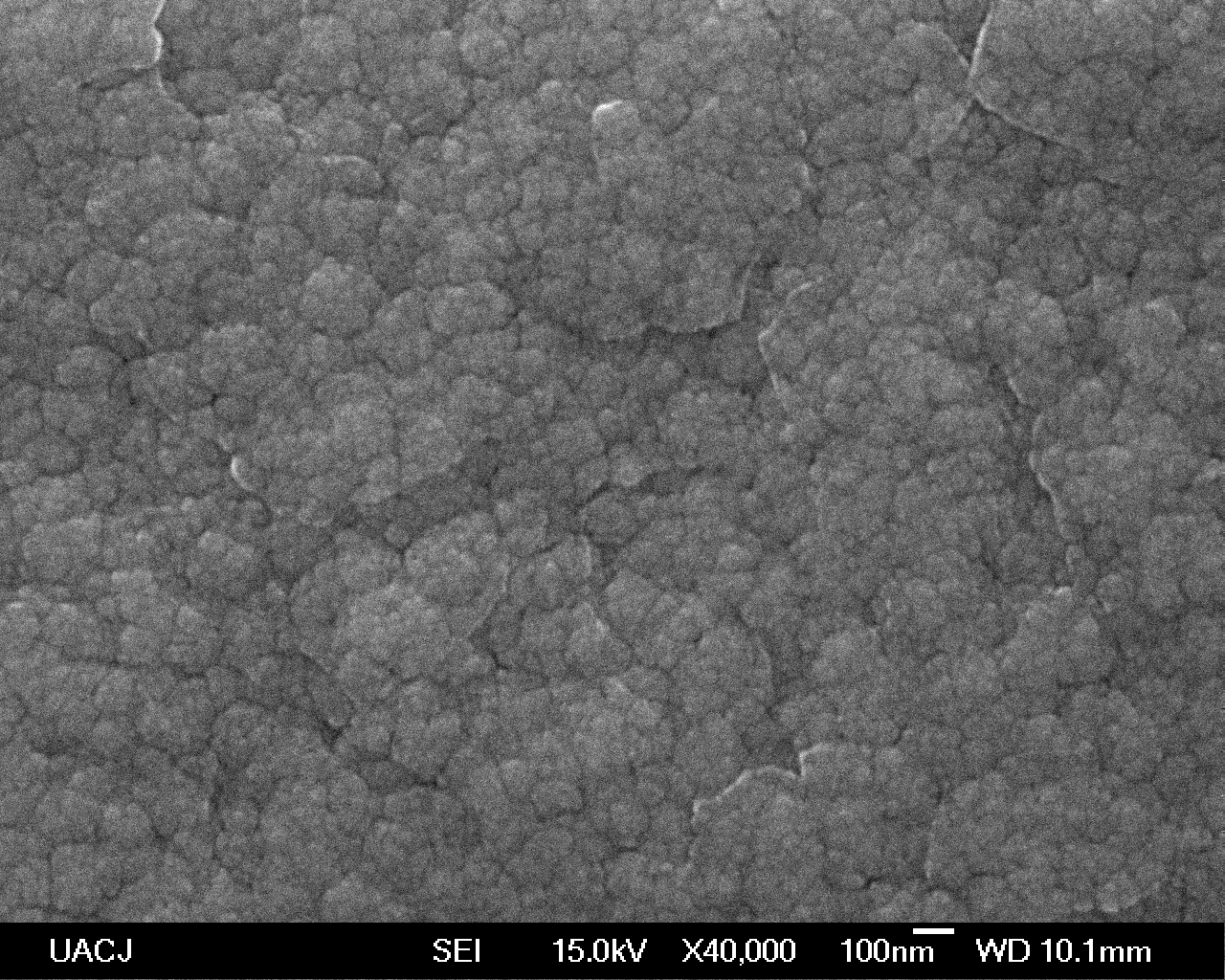}
\includegraphics[width=4cm]{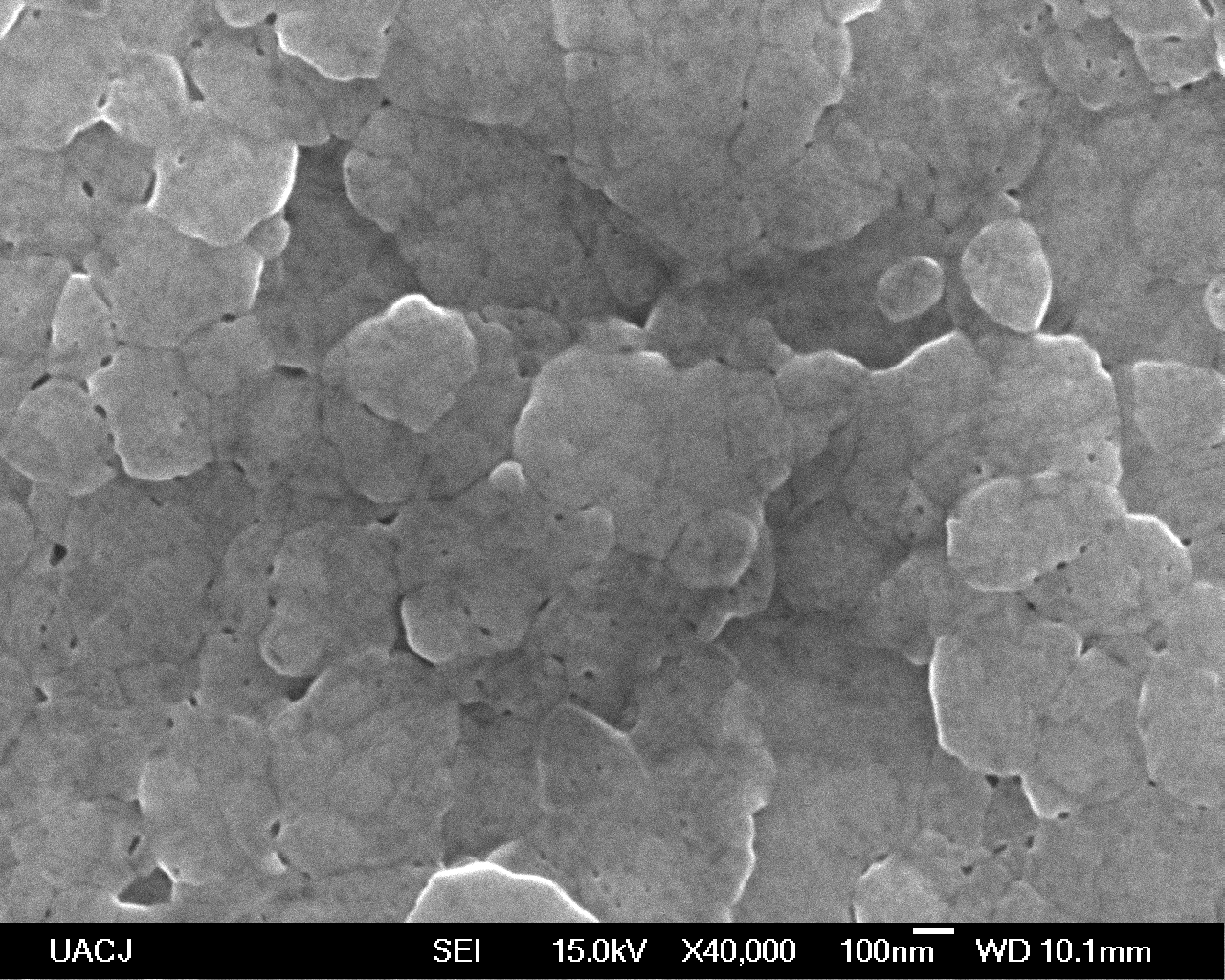} \includegraphics[width=4cm]{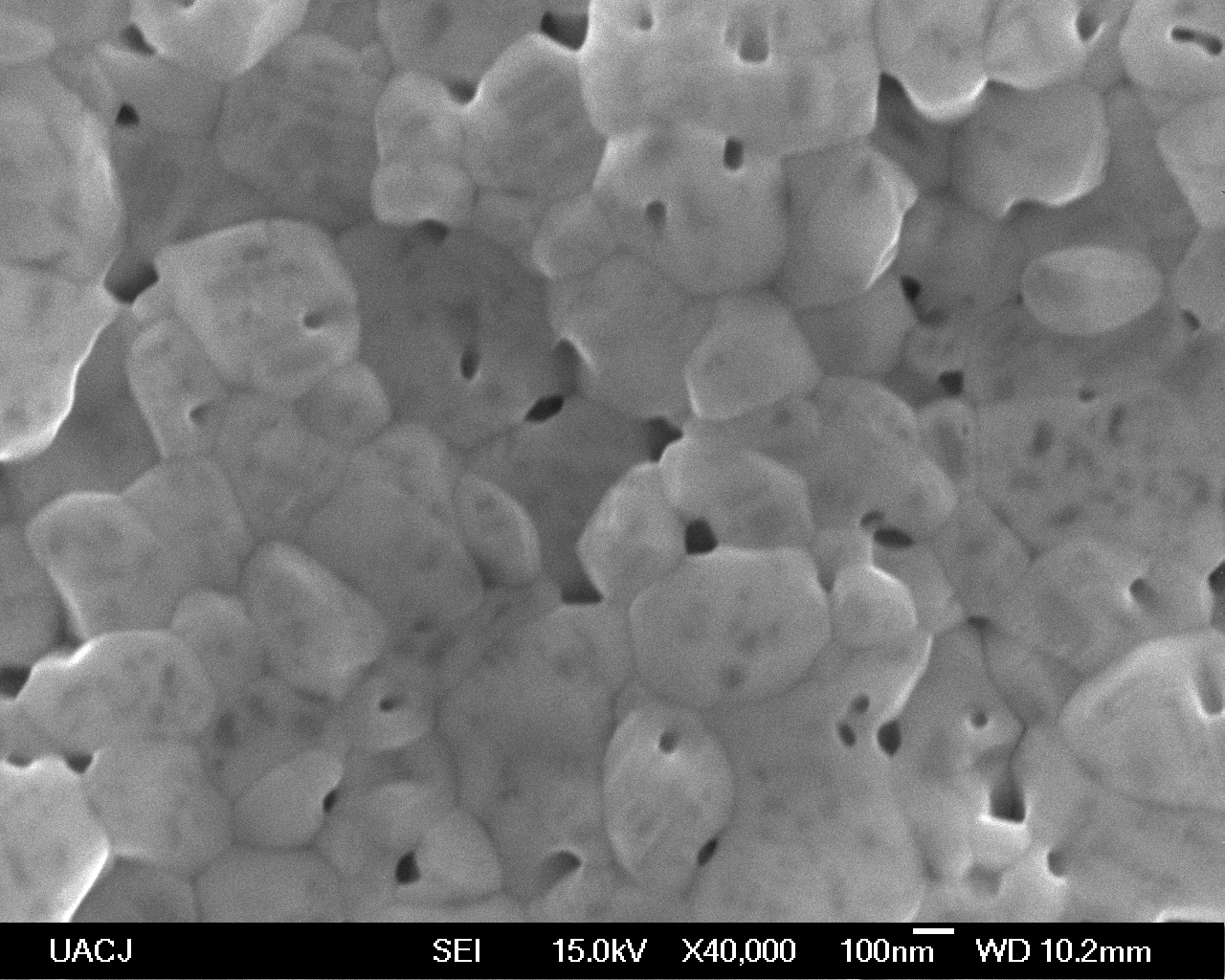}
\caption{From left to right and top to bottom, SEM micrographs for the as-deposited film F0, and the films F575, F675, F775, F875, and F1000.}
\label{ss}
\end{center}
\end{figure}

The micrographs in Fig. \ref{ss} show the evolution of the morphology for the as-deposited film and the films exposed to heat treatments.
\begin{table}[b!]
  \centering 
  \caption{Annealing temperature $T_{ann}$, film thickness $t$, deposition rate $D_r$, optical band gap $E_g$, and oxygen to tantalum ratio for the films.}
  \label{details}
  \begin{tabular}{c|c|c|c|c|c}
\hline
Film   & $\frac{T_{ann}}{^\circ \; \text{C}}$ & $\frac{t}{\mu\text{m}}$  & $\frac{D_r}{\text{\AA} \cdot \text{s}^{-1}}$ & $\frac{E_g}{eV}$ & $\frac{\text{O at\%}}{\text{Ta at\%}}$   \\ \hline
 F0  & --- & 2.4 & 2.8 & --- &  \\
 F575  & 575 & 2.4 & 2.8 & 2.4 & 2.6 \\
  F675 & 675& 2.4& 2.8 &3.5  & 2.7 \\
  F775 & 775 & 2.4 & 2.8 &3.5 & 2.6 \\
  F875 & 875 & 2.4 & 3.6 &3.6 & 2.5  \\
    F1000  & 1000& 2.4 &3.6 & 3.8 & 2.8  \\
\hline
\end{tabular}
\end{table}
Overall, the micrographs exhibit a flake-like microstructure. The film F575 shows a slight change in the morphology with respect to the as-deposited sample, for round and small plates can be distinguished, revealing the appearance of discrete grain boundaries. As the temperature is increased beyond 575$^\circ$C, larger plates formed reflecting surface fracture effects due to both grain formation and thermal expansion occurring in deeper layers. At the same time, small plates start to fuse among each other. Further increase of the temperature above 775$^\circ$C transforms the flake-like microstructure into more homogenous plates, suggesting strong diffusion effects. At 1000$^\circ$C the film develops into a granular structure with a large amount of voids. 
\begin{figure}[t!]
\begin{center}
\includegraphics[width=6cm]{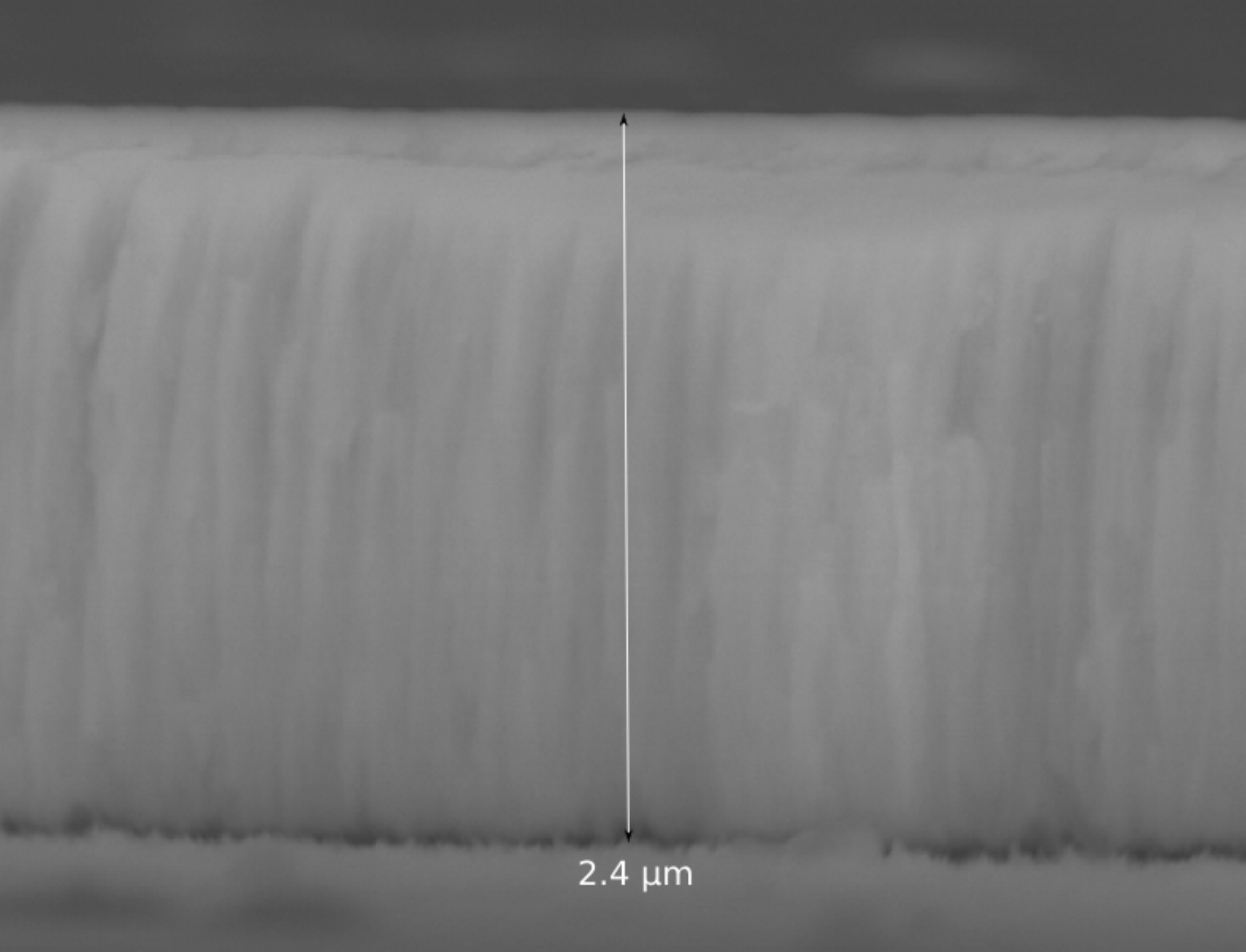}
\includegraphics[width=6cm]{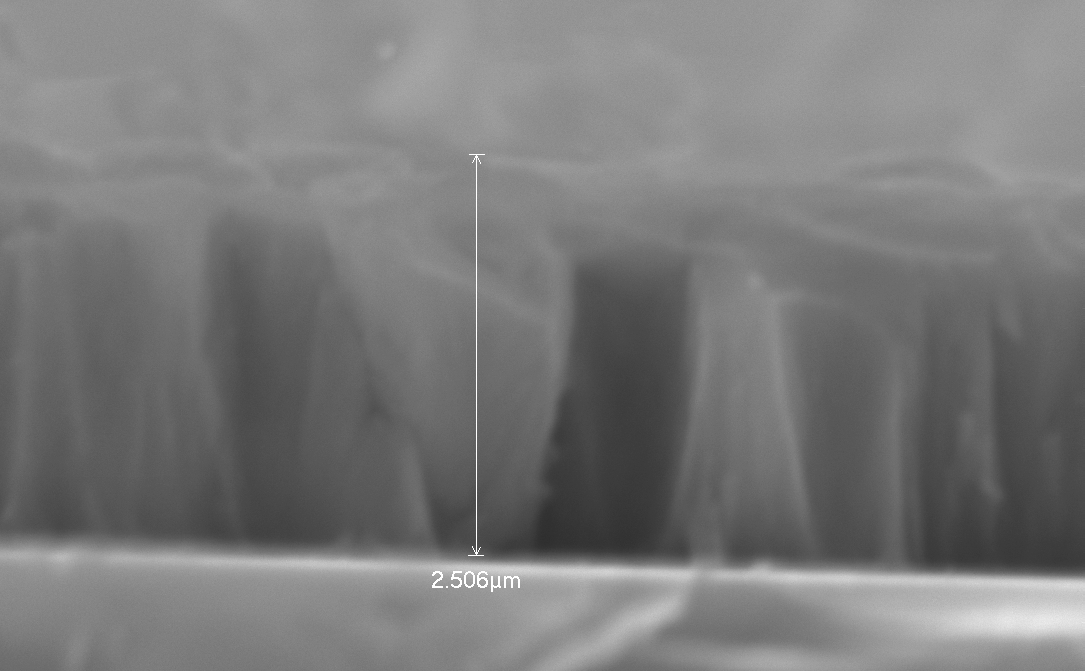}
\includegraphics[width=6cm]{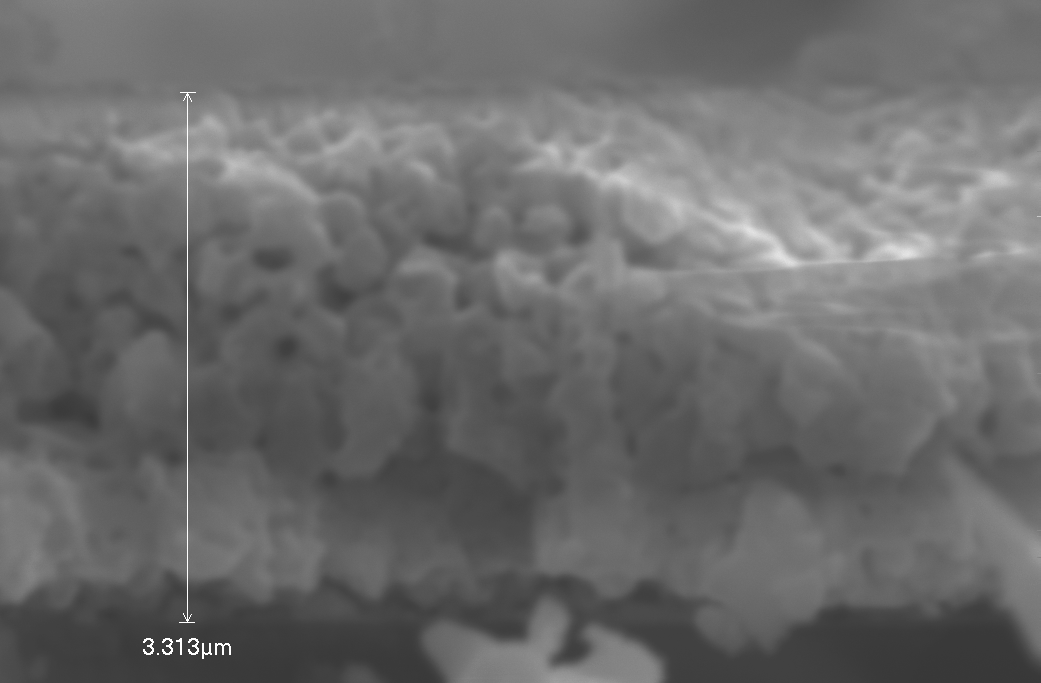}
\caption{Top:  Film F0 showing smooth and fine growth. Middle: Film F575 showing similar microstructure as film F0 and a slight increase in thickness. Bottom: Film F1000 develops into large and disordered grains and a considerable increase in thickness.}
\label{s11Canto}
\end{center}
\end{figure}
In this same sample, some dark spots on the grains are easily visible which can be even identified from the film F875. To elucidate the nature of these spots, we took backscattering electron images (not shown). The results suggest that the spots belong to another oxide phase of Ta, most probably the hexagonal phase of Ta$_2$O$_5$ or a suboxide (TaO$_x$). An EDS (Energy dispersive x-ray spectroscopy) analysis performed on these spots did not observe significant difference in the atomic concentration of the species. However, using XPS we computed the stoichiometry of the films as the atomic concentration ratio of oxygen to tantalum (O:Ta) and found values close to stoichiometry, 2.5; showing that the films are fully oxidized (see Table \ref{details}). In regards to the grain size, according to the SEM image of F1000, the analysis indicates that the diameter ranges from 100 nm to 500 nm with an average size of 300 nm. 

In order to verify a possible thermal expansion of the films, we have taken cross-sectional micrographs for the films F0, F575, and F1000 (Fig. \ref{s11Canto}). In the images we notice that the thickness for F0 is about 2.4 $\mu$m and the texture is smooth and homogeneous. A similar situation occurs for the film F575 although we observe a slight increase in the thickness (2.5 $\mu$m.) The film F1000 develops into large disordered grains, leaving large voids between them and a thickness of 3.3 $\mu$m; representing an increase of 900 nm with respect to the as-deposited film. This unambigously demonstrates that there is a considerable expansion effect as the films oxidize. To the best of our knowledge, there is no other work reporting on this matter, we think that this a revelevant finding when manufacturing films with the method used here and hence one can infere that this effect is also present in thin films.

\subsection{Crystalline Structure}

\begin{figure}[t!]
\begin{center}
\includegraphics[width=9.5cm]{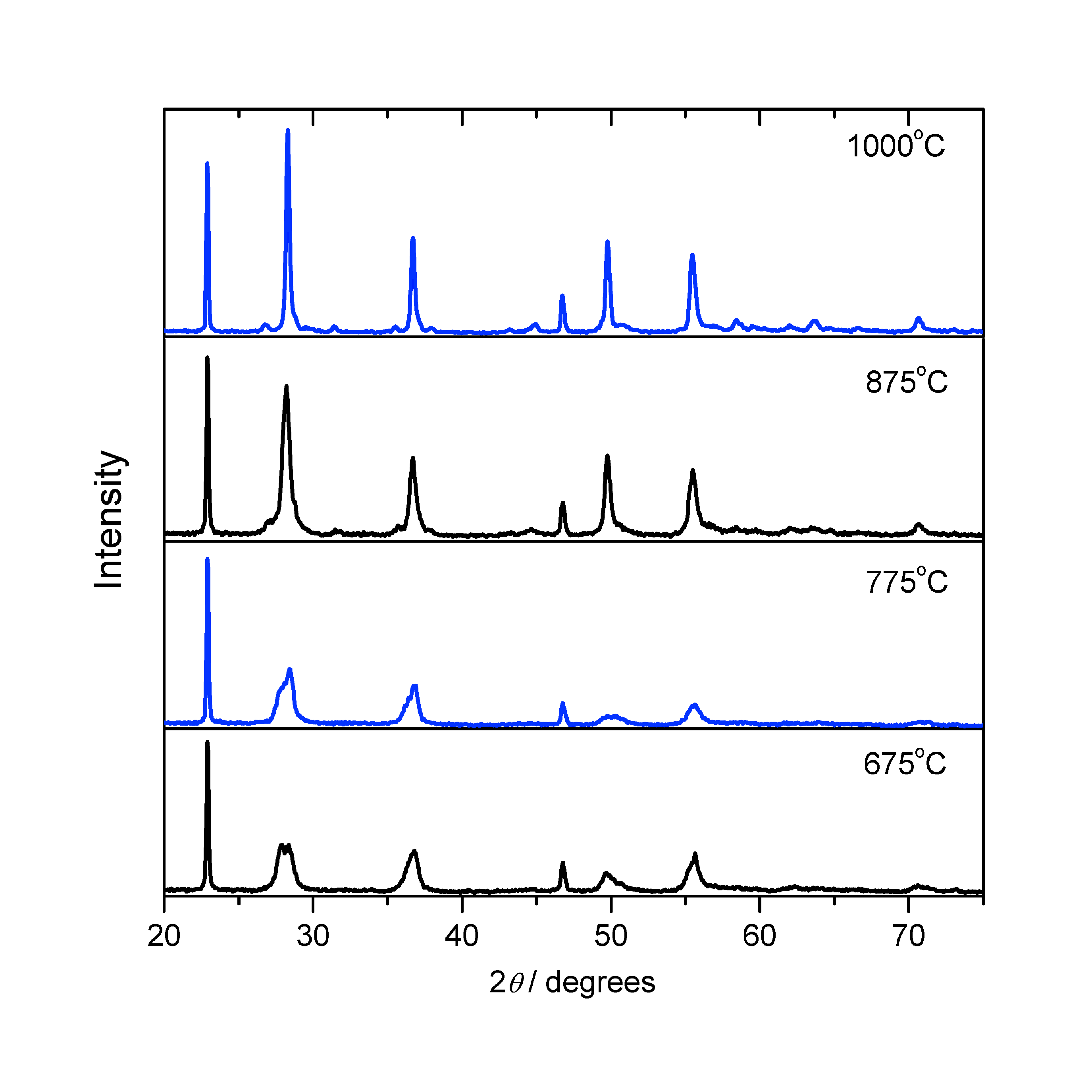}
\caption{X-ray diffraction patterns for the crystalline films F675-F1000.}
\label{xrdS6}
\end{center}
\end{figure}

\begin{figure}[t!]
\begin{center}
\includegraphics[width=9.5cm]{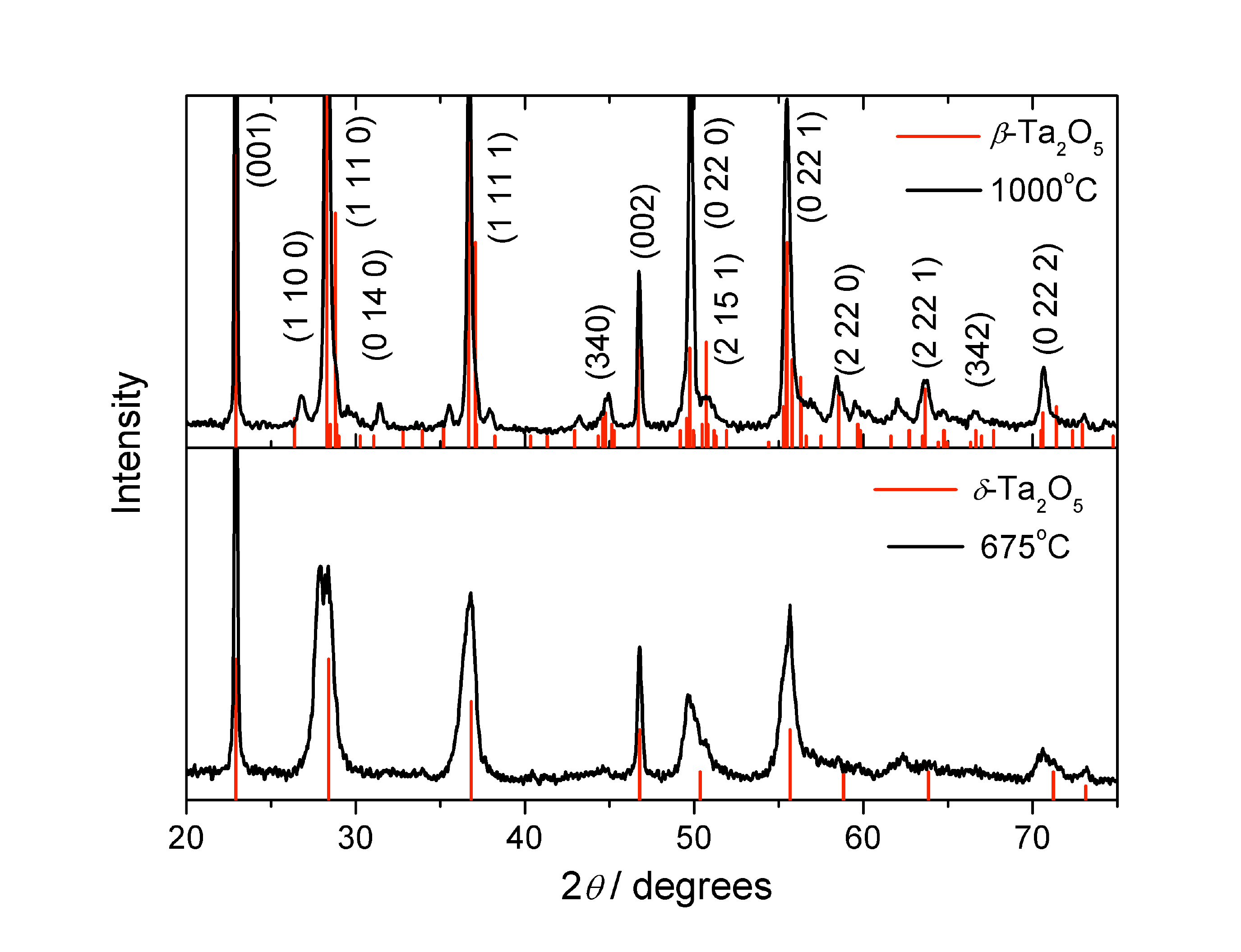}
\caption{X-ray diffraction patterns for the film F1000 (top) and the film F675 (bottom) in comparison to the reference patterns PDF 025-2209 and PDF 019-1299, respectively.}
\label{xrds16vsref}
\end{center}
\end{figure}

The as-deposited film as well as the film F575 do not display any feature in the XRD pattern, indicating that the atomic structure remains amorphous and hence the corresponding XRD patterns are not shown here. On the contrary, the films F675-F1000 display at least six prominent peaks at angles around $2\theta=$(22.8, 28.3, 36.7, 46.7, 49.7, and 55.8)$^\circ$ which point to the formation of crystalline structures (see Fig. \ref{xrdS6}).
These patterns can be indexed to either the orthorhombic phase $\beta-$Ta$_2$O$_5$ (PDF 00-025-0922 with lattice parameters $a=6.1980$ \AA, $b=40.2900$ \AA, $c=3.8880$ \AA, and $\alpha=\beta=\gamma=90^\circ$ for the spatial group $P2_12_12$; or PDF 01-089-2843 with lattice parameters $a=6.2000$ \AA, $b=3.6600$ \AA, $c=3.8900$ \AA, and $\alpha=\beta=\gamma=90^\circ$ for the spatial group $Amm2$) or the hexagonal phase $\delta-$Ta$_2$O$_5$ (PDF 00-019-1299 with lattice parameters $a=b=3.6240$ \AA, $c=3.8800$ \AA, $\alpha=\beta=90^\circ$, and $\gamma=120^\circ$, with presumably spatial group $P6/mmm$ \cite{afukumoto97a,ynwu11a}). As of now, there is still an ongoing debate regarding the specific space group of these two phases \cite{spwalton16a,jykim14a,shlee13a,jlee14a,zhelali14a,yguo15a,jykim15a}. According to our results, it is clear that as the annealing temperature is raised more peaks in the diffraction patterns show up. In the coating F1000 we identified at least 11 evident peaks and 14 peaks of higher orders with low intensity. Film F875 displays at least 9 peaks whereas the films F575 and F675 both exhibit 7 peaks. The position of the peaks for all samples remain around the same angles, an indicative that there is no influence of the annealing temperature on the cell size. The pattern for the film F875 is quite similar to that of film F1000 and one can consider that they have the same crystalline phase. Similarly, the patterns for films F575 and F675 have the same shape and one can assume they have the same crystalline structure. To identify the phase of our films, we plotted only the patterns of the films F1000 and F675 against the reference patterns PDF 00-025-0922 and PDF 00-019-1299, respectively (Fig. \ref{xrds16vsref}). By comparison we can observe that the pattern of the $\delta$ phase agrees well with our measurements, however it is not  capable of reproducing many features exhibited by the pattern of the film F1000, especially the low intensity peaks scattered along the whole spectrum. By contrast, it is noticeable that the $\beta$ phase remarkably matches the pattern of the sample F1000. The pattern PDF 01-089-2843 also reproduces the main features of F1000 but does not match the minor peaks, especially those appearing beyond $2\theta=55^\circ$. Due to the overlapping of the most instense peaks of both reference patterns shown in Fig. \ref{xrds16vsref} we can not rule out the coexistence of both phases in the films.
In any case, the existence of additional peaks for the films annealed at high temperature can signify a transition in the crystalline structure. Since the pattern of the $\beta$ phase is in much better agreement with the pattern of F1000 than  the pattern of the $\delta$ phase is, we thus presume that for annealing temperatures above 775$^\circ$C the orthorhombic phase dominates the system. Hence, the patterns for F875 and F1000 can be indexed to the $\beta$ phase whereas the diffraction patterns for the films F675 and F775 can be indexed to the $\delta$ one.

We further note from Figs. \ref{xrdS6} and \ref{xrds16vsref} that there is peak sharpening as the temperature is raised suggesting a change in the crystallite size. S.-J. J. Wu et al. and Dimitrova et al. \cite{tdimitrova01a,sjjwu09a} prepared a series of Ta$_2$O$_5$-thin films by the RF magnetron sputtering technique under different oxygen concentrations and annealing temperatures (700$^\circ$C-850$^\circ$C). They attributed the peak sharpening to both the relatively high annealing temperature  and high oxygen concentration. We also observe peak sharpening but since all films were grown in air, this cannot be attributed to high oxygen concentration; rather, our results lend weight to the conviction that the annealing temperature has a major impact on the oxidation process. The reason for this apparent contradiction may lie in the fact that at low deposition rates (as those used by Wu and Dimitrova et al.: 0.4 \AA/s and 0.8 \AA/s, respectively.) the tantalum atoms have enough time to create bonds with both the residual oxygen atoms present in the vacuum chamber and the injected oxygen and thus create tantalum suboxides in lower layers, while the upper layer is fully oxidized. This may explain the minor influence of the oxygen flow on the crystallinity of the films \cite{xmwu93a}.

To estimate the crystallite size $D$ we used the well-known Scherrer equation
\begin{equation}
\label{sch}
D=\frac{K\lambda}{\Gamma \cos\theta},
\end{equation}
where $K=0.9$, $\lambda=1.5406\;\text{\AA}$  and $\Gamma$ is the full width at half maximum of a given peak. Accordingly, taking the peak at $2\theta=28.3^\circ$ for the four films F675-F1000, we found that $\Gamma=$(0.79, 0.69, 0.52, 0.27)$^\circ$ and hence $D=$(10, 12, 16, 30)nm, respectively. This reaffirms the nanometric microstructure of the films and signifies that the crystallite size increases as function of the annealing temperature. 

\subsection{Vibrational modes: Raman spectroscopy}
The vibrational modes of the samples were examined by Raman spectroscopy. For the purposes of this work, we used as a reference the Raman spectrum of Ta$_2$O$_5$ powder. The reference sample is a commercial powder acquired from Alfa Aesar with purity of 99.99\%. 
\begin{figure}[b!]
\begin{center}
\includegraphics[width=10cm]{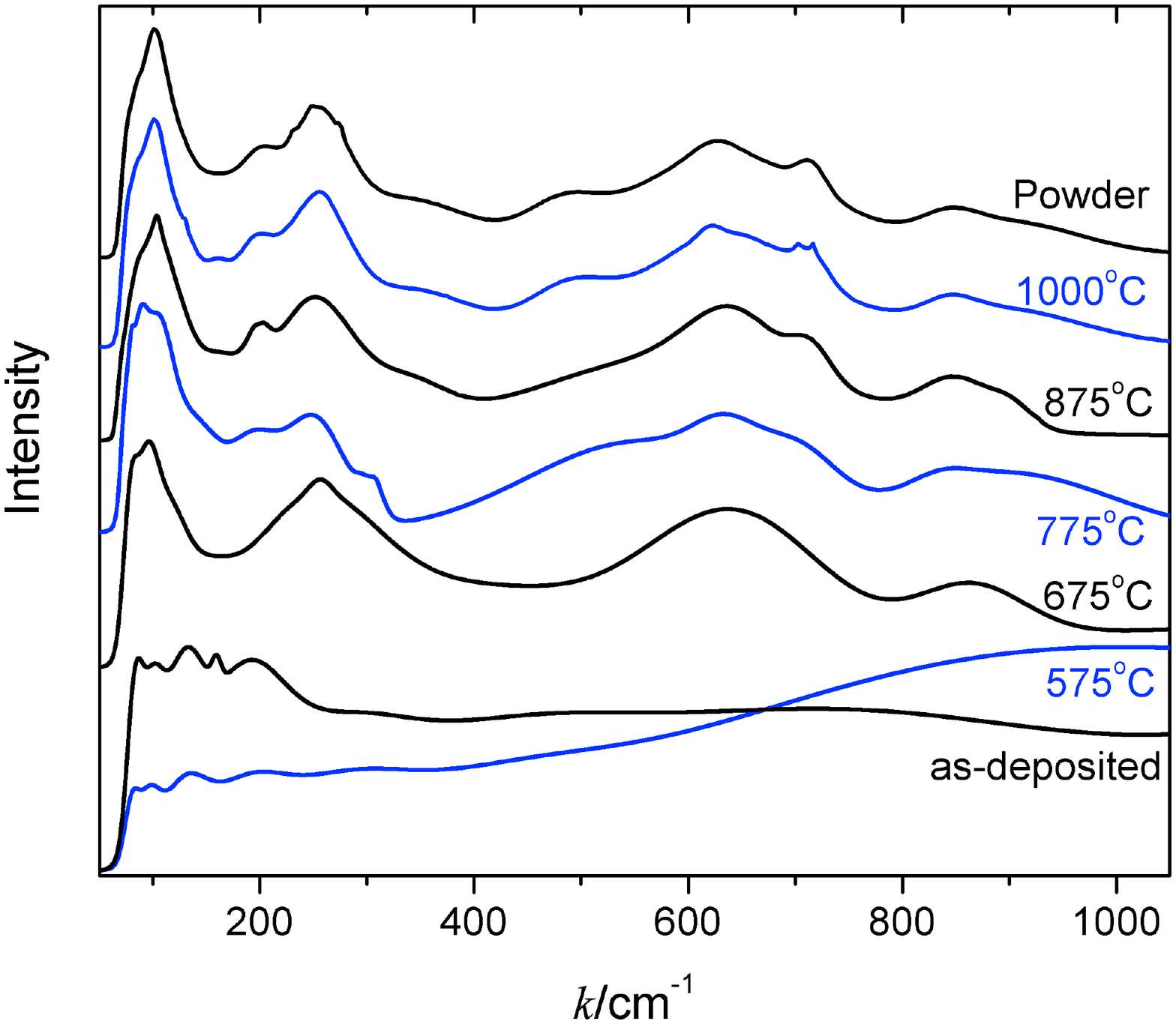}
\caption{Raman spectra for films annealed at different temperatures.}
\label{ramans}
\end{center}
\end{figure}
To eliminate humidity and ensure the crystallinity, the powder was annealed at 1000$^\circ$C for 1 h in air.  Figure \ref{ramans} shows the Raman spectra of the films in comparison to the reference spectrum. The latter agrees quite well with the spectrum reported in the literature \cite{yzhu05a,rrkrishnan09a,cjoseph12a}. Starting with the Raman spectrum of the as-deposited film we note that it does not display characteristic peaks but only typical background due to fluorescence. This is expected since Ta (amorphous or crystalline) does not exhibit characteristic peaks. Similarly, the spectrum of F575 does not display any characteristic peaks, however, we do observe that there is a change in the shape of the background suggesting that the film, although still Ta amorphous, is undergoing a gradual oxidation process. This is confirmed as the annealing temperature is raised.  The spectrum of the film F675 clearly starts to resemble the reference spectrum although minor peaks are not present. 

According to the analysis done in the previous section, we can attribute the shape of this spectrum to the $\delta$ phase, for if we analyze the spectra of the films F675-F1000 we notice that those minor features appear as the transition from the $\delta$ phase to the $\beta$ phase takes place. By comparing the reference spectrum with the spectrum of F1000, we see that they are almost identical indicating that the $\beta$ phase has been achieved.

Joseph et al. \cite{cjoseph12a} proposed that $\beta-$Ta$_2$O$_5$ has 19 active vibrational modes. The number of modes was set based on the number of peaks reproduced by fitting a reference experimental spectrum with a fitting software. 
\begin{figure}[b!]
\begin{center}
\includegraphics[width=10cm]{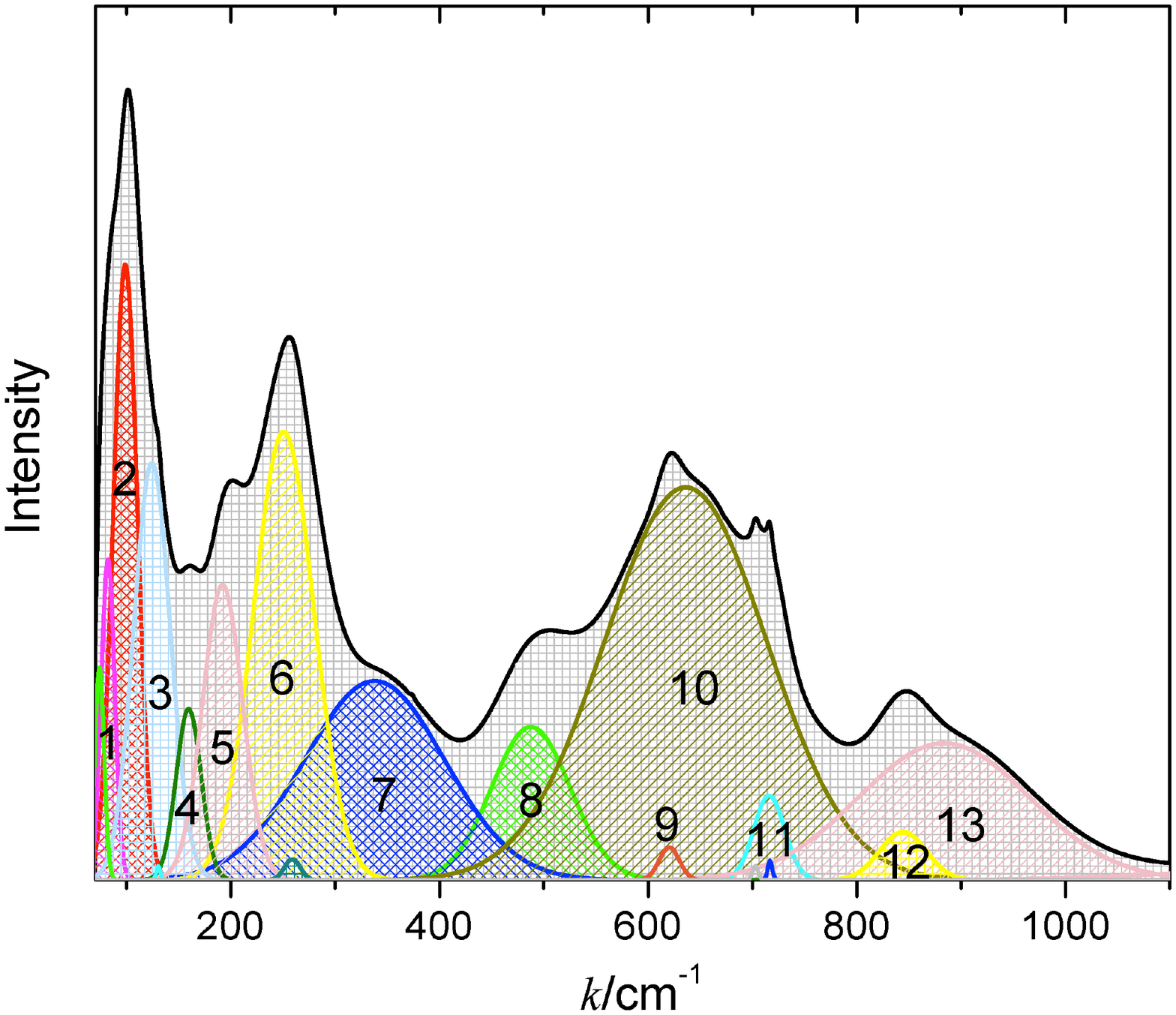}
\caption{Deconvolution of the Raman spectrum for the film F1000.}
\label{ramandeconv}
\end{center}
\end{figure}
\begin{table}[t!]
  \centering 
  \caption{Peak parameters according to the deconvolution realized on the Raman spectrum of the film F1000. The numbers in parenthesis are the values reported by Joseph et al. \cite{cjoseph12a}.}
  \label{ramandata}
  \begin{tabular}{c|c|c}
\hline
Band & Frequency/cm$^{-1}$ & FWHM/cm$^{-1}$ \\ \hline
 1  & 82 (78)  & 14.8 \\
  2 &  99 (106) & 26.9 \\
  3 &  124 (139)  & 41.0\\
  4 &  160  (---) & 28.0 \\
   5  &  192 (196)  & 44.8 \\
   6  & 251(245)  & 67.9 \\
  7 & 338 (338) & 157.2 \\
  8 &  488 (494) & 88.4 \\
  9 & 620 (612) &  22.3\\
  10 &  635 (642) & 183.6 \\
   11  & 716 (711)  & 34.7  \\
12 & 844 (844) &  52.4\\
13 & 883 (903) &  192.2\\
\hline
\end{tabular}
\end{table}
However, the number of peaks strongly depends on the type of function used during the fitting procedure. To determine the peak parameters, we took the spectrum of film F1000 and perform a fitting procedure similar to that conducted by Joseph et al. The deconvolution of the spectrum into Gaussian peaks gave us the number of peaks and the peak parameters (see Fig. \ref{ramandeconv}). 

Table \ref{ramandata} gives the results which are in good agreement with those reported by Joseph et al. The modes for F675 appear at (82, 99, 251, 635, and 844)cm$^{-1}$ and can be attributed to the $\delta$ phase for as the annealing temperature is incremented, more features appear resembling the reference spectrum. From the spectrum of the film F675 we can see emerging features which are greatly enhanced for F775, F875, and finally for F1000. According to previous work \cite{cjoseph12a,psdobal00a}, bands 1-3 can be assigned to lattice modes, while bands 4-7 are due to bending modes $\delta$(Ta-O-Ta) and $\delta$(O-3Ta). The region between 400 cm$^{-1}$ and 1000 cm$^{-1}$ is where the stretching modes are found; bands 8-11 are attributed to $\nu$(O-3Ta) whereas bands 12-13 are related to $\nu$(Ta-O-Ta) \cite{hono01a, wandreoni10a}. Bands below 100 cm$^{-1}$ belong to interactions of different Ta polyhedra. Balachandran et al. observed a band at 627 cm$^{-1}$ which they ascribed to the vibrations in the TaO6 octahedra \cite{ubalachandran82a}. In our spectrum we observed a peak at 635 cm$^{-1}$ (band 10) that can be associated to the observations reported by Balachandran et al. The peak at 844 cm$^{-1}$ (band 12) suggests the presence of TaO5 (pentagonally coordinated groups), while a peak near band 13 was spotted by F. L. Galeener et al. in Ta$_2$O$_5$ deposited by rf magnetron sputtering \cite{flgaleener80a}. 

According to the XRD analysis, the modes (124, 160, 192, 338, 488, 620, 716, and 883)cm$^{-1}$, can be associated to the orthorhombic phase. Thus the foregoing analysis reinforce the view that a continuous amorphous-to-crystal transition is gradually occurring when the annealing temperature is increased beyond a threshold temperature between 575$^\circ$C and 675$^\circ$C. Also there seems to be a crystalline transition from hexagonal to the orthorhombic symmetry for annealing temperatures beyond 775$^\circ$C. We observe that this transition can be characterized by Raman spectroscopy. Furthermore, contrary to Joseph et al. who reported the Raman spectrum of amorphous Ta$_2$O$_5$, we do not observe evidence supporting the presence of this phase. From these findings, it is clear that there is a straightforward transition from tantalum amorphous to crystalline Ta$_2$O$_5$.

\subsection{Optical properties: UV-VIS}

\begin{figure}[t!]
\begin{center}
\includegraphics[width=9.5cm]{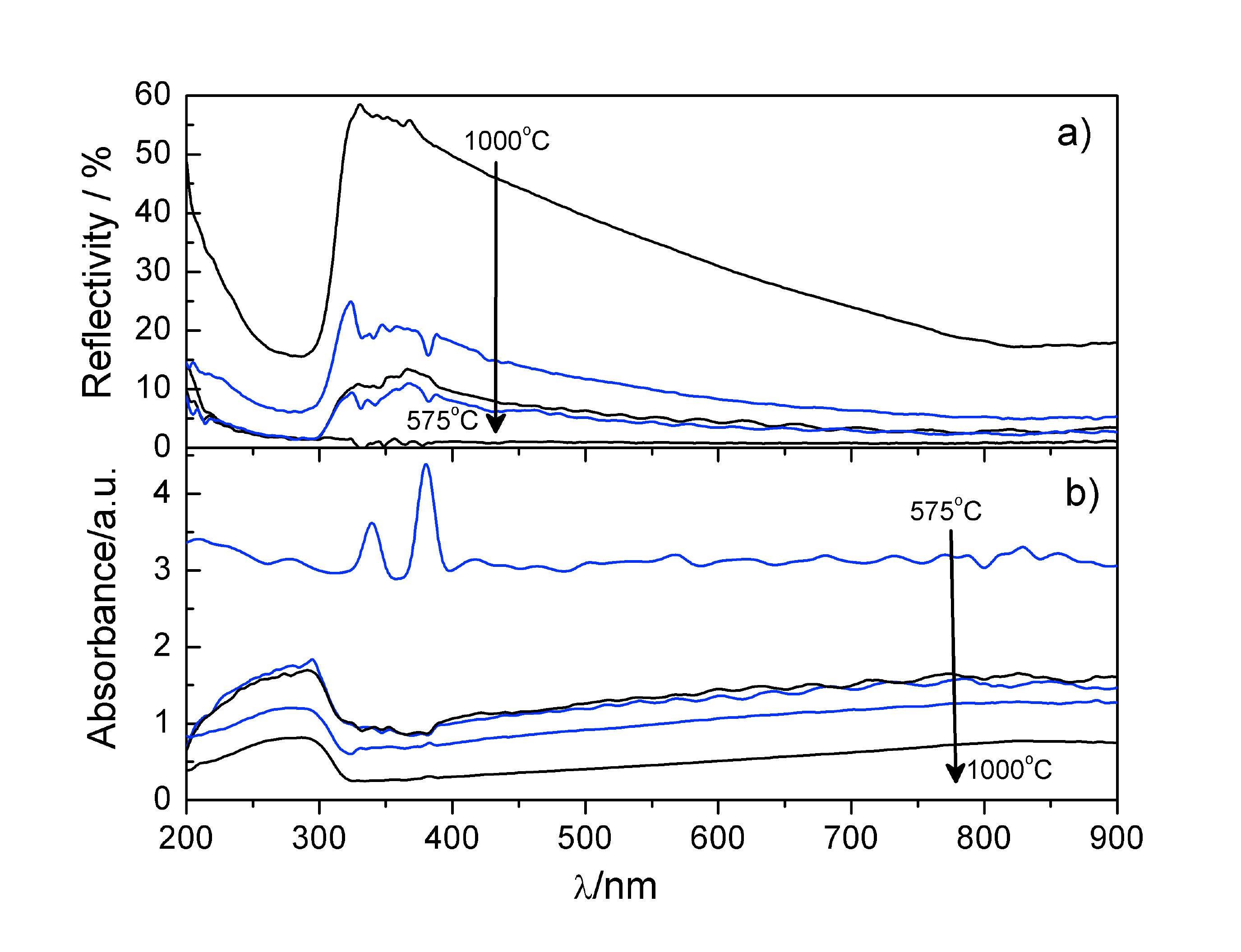}
\caption{Absorbance (a) and reflectivity (b) for the annealed films. The arrows indicate the direction in the variation of the annealing temperature.}
\label{opticalprop}
\end{center}
\end{figure}

As a semiconductor, crystalline Ta$_2$O$_5$ possess an optical band gap whose energy ranges from 3.5 eV to 4.5 eV \cite{rrkrishnan09a,yhpai08a,mstodolny09a,slin13a,hyu13a}. It is therefore of prime importance to measure this property taking as parameter the annealing temperature. For this purpose we carried out optical measurements using an UV-VIS system. We first measured the dependence of the reflectivity ($R$) and the absorbance on the wavelength ($\lambda$) by the method of diffuse reflectance (see Fig. \ref{opticalprop}). Starting with the as-deposited film we visually observed a quite smooth and polished surface. This implies that no diffuse reflection is generated and we do not report this result. As seen from the micrograph of the film F575, there is no considerable change in the morphology with respect to the as-deposited film. 
One can visually observe that it remains of the same optical grade as when deposited and thus one would expect specular reflection. This explains the negligible reflection spectrum of F575 shown in the Fig. \ref{opticalprop}(a). The case for the films with annealing temperatures $\ge$ 675$^\circ$C is radically different for they slowly turn into a white and rough flaking structure as we evidenced from the SEM analysis, resembling the white granular morphology of the reference powder and diffuse reflection dominates. From Fig. \ref{opticalprop}(a) we see that the curves shift to higher reflection percentages as the system further crystallizes. A similar trend is observed in the absorption plot shown in Fig. \ref{opticalprop}(b).

\begin{figure}[t!]
\begin{center}
\includegraphics[width=9.5cm]{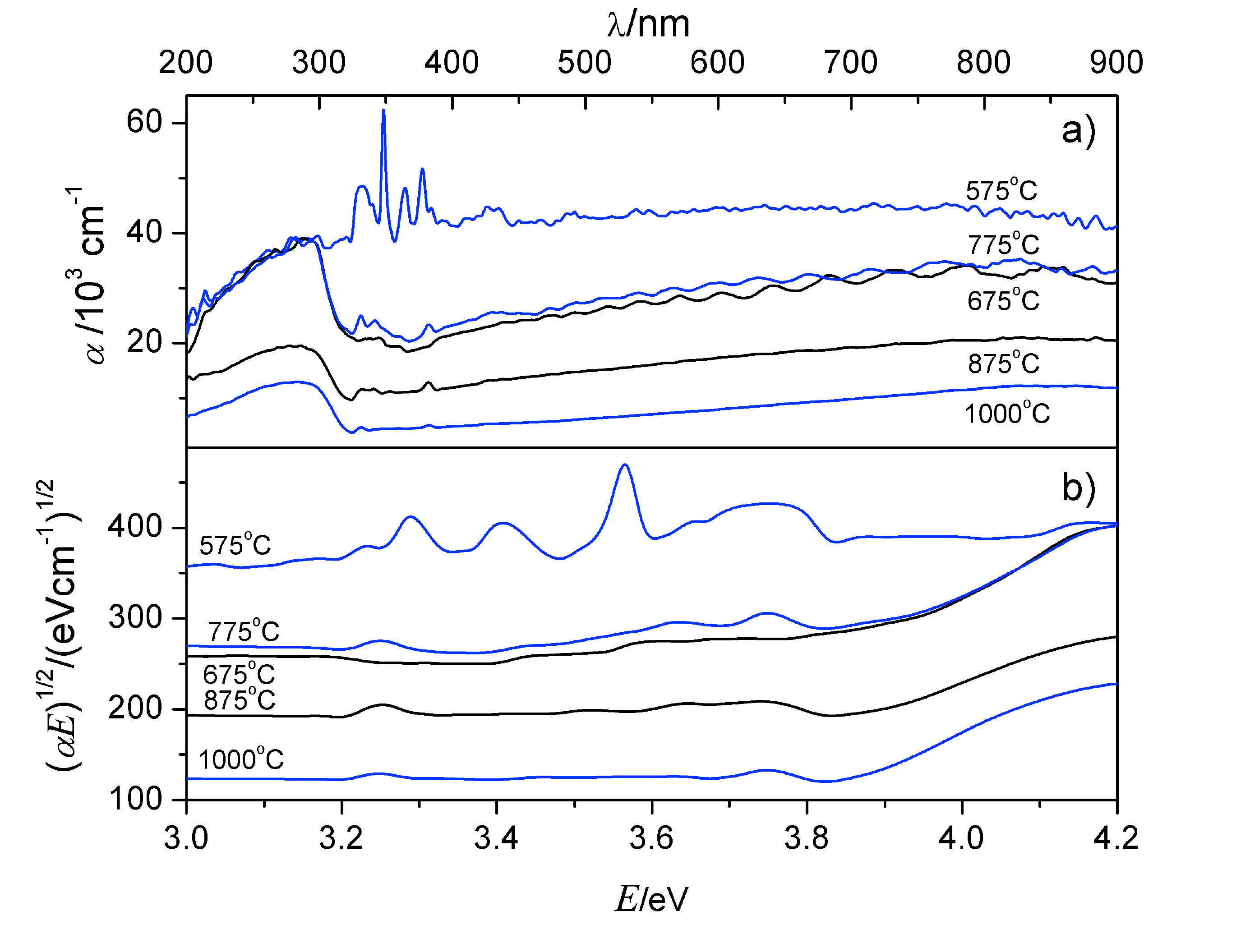}
\caption{Absorption coefficient (a) and Tauc's plot (b) for the annealed films.}
\label{absorptionCoef}
\end{center}
\end{figure}

The optical absorption coefficient $\alpha$ can be determined from the measurements of $R(\lambda)$ using the Kubelka-Munk equation:
\begin{equation}
\label{alphacoef}
\alpha=s\frac{(1-R)^2}{2R},
\end{equation}
where $s$ is the scattering factor. The absorption coefficient is plotted in Fig. \ref{absorptionCoef} (a) where we observe the same trend as in the absorption plot. 
This coefficient is related to the incident photon energy $E$ by Tauc's relationship \cite{jtauc66a}: 
\begin{equation}
\label{tauceq}
\alpha E=\alpha_0(E-E_g)^n,
\end{equation}
here $\alpha_0$ is an energy independent constant, sometimes called the band tailing parameter; $E_g$ is the optical energy gap; and $n$ is called the power factor of the transition mode. The latter depends upon the atomic order of the material and the photon transition. The possible values of $n$ are 1/ 2, 2, 3/2, and 3; the value adopted depends on the photon transition, i.e., direct allowed, indirect allowed, direct forbidden and indirect forbidden transition, respectively. To determine the most suitable value of $n$, we plotted all possible values of $n$ against the photon energy and found that the most adequate value is $n=2$.

To estimate the optical gap, we plotted $(\alpha E)^{1/2}$ versus $E$ using the data obtained from the optical absorption spectra [see Fig. \ref{absorptionCoef}(b)]. The curves indicate that the obtained plotting gives a straight line in a certain region. To obtain the magnitude of the optical gap, we extended this straight line to intercept the $E$-axis at $(\alpha E)^2=0$. With this method we found the values tabulated in Table \ref{details}. We observe that there is an increase in $E_g$ as function of temperature which is accompanied with the crystallization process. The magnitude of the energy gap gradually increases from ($2.4\pm 0.2$)eV in F675 to ($3.8\pm 0.2$)eV in F1000, the latter value in good agreement (within in the limits of error) with other reports \cite{rrkrishnan09a,yhpai08a,mstodolny09a,slin13a,hyu13a}. We would like to remark that the films F675 and F775, which are related to the hexagonal phase, show band-gap magnitudes (3.6 eV) slightly smaller than the values found for the films F875 and F1000 which are associated to the orthorhombic phase. To the best of our knowledge there is no experimental report in the literature on the magnitude of the band gap for the hexagonal phase. We think that this is an interesting finding that could be valuable for future reference in theoretical and experimental investigations.

\section{Conclusions}
We investigated the dependence of the crystalline and physical properties on annealing temperature in tantalum-pentoxide films grown by radio frequency magnetron sputtering. The annealing temperature used in this work ranges from 575$^\circ$C to 1000$^\circ$C. We performed X-ray diffraction, scanning electron microscopy, Raman spectroscopy, and UV-VIS spectroscopy to monitor the evolution of the crystalline structure, morphology, vibrational modes, and optical properties, respectively. Our results indicate that upon annealing the films undergo a crystallization above 675$^\circ$C, followed by a possible structural transition beyond 775$^\circ$C. The crystallite size slowly increases as the temperature increases and the number of vibrational modes is affected suggesting that indeed crystallization occurs. We also determined that Raman spectroscopy is sensitive to the structural transition and it can be used to characterize the crystalline phases of Ta$_2$O$_5$ by analyzing the number of modes. Finally, we studied the optical properties and estimated the optical band gap. We found that the gap magnitude is greatly influenced by the annealing temperature with values ranging from 2.4 eV to 3.8 eV, the latter in better agreement with the values reported in the literature. We further realized that the values for hexagonal Ta$_2$O$_5$ are slighly smaller than the values for the orthorhombic phase. 

\section*{Acknowledgements}
We are grateful to Jorge Ivan Betance, Daniel Aguilar, and Wilian Cauich for their technical support during the SEM, DRX, and XPS sessions, respectively. The authors gratefully acknowledge the support from the National Council of Science and Technology (CONACYT) Mexico. C. I. Rodr\'iguez acknowledges the support from the laboratory of material analysis at the Universidad Tecnol\'ogica de Ciudad Ju\'arez.

\end{document}